\documentstyle[12pt]{article}
\advance \textheight by \topskip
\begin{document}
\begin{titlepage}
\hbox to \hsize{}
\hbox to \hsize{\hfil hep-ph/9610249 }
\hbox to \hsize{\hfil Revised }
\vfill
\large \bf
\begin{center}
On the extreme behaviour of $g_1(x,Q^2)$ at $x\to 0$.
\end{center}
\vskip 1cm
\normalsize
\begin{center}
{\bf S. M. Troshin}\\[1ex]
{\small  \it Institute for High Energy Physics,\\
Protvino, Moscow Region, 142284 Russia}
\end{center}
\vskip 1.5cm
\begin{abstract}
On the basis of the $U$--matrix form of $s$--channel unitarization, we
consider constraints unitarity provides for the spin structure function
$g_1(x,Q^2)$ at $x\to 0$. Corresponding constraint for the spin
structure function $h_1(x, Q^2)$ is given along.\\ [2ex]
\end{abstract}
\vfill
\end{titlepage}

The problem of spin structure of a nucleon during the last decade
has become a very significant one. Experiments in this field continue
to bring new facts which can change earlier interpretations.
As it appears now, the behaviour of the function $g_1$ at
$x\to 0$ is becoming  crucial in the evaluation
of the total hadron helicity carried by  quarks.
Recent experimental results indicate that the function
$g_1(x)$ might have a rising behaviour at $x\to 0$
 \cite{mal} contrasting to a smooth Regge dependence
$g_1\sim x^{-\alpha_{a_1}}$ (with $-0.5<\alpha_{a_1}<0)$
used in the experimental analysis. Indeed, the  dependence
$g_1^n(x)=-0.2/x^{0.8}$ observed at small $x$ in the experiment
E154 at SLAC \cite{mal} when taken as an extrapolation to
$x=0$ could significantly change the first moment of the
structure function $g_1$. Thus, the problem of small-$x$
seems to be important in both cases of unpolarized and polarized
DIS.

It has been discussed in several papers basing on some general
considerations \cite{close}, DLA in QCD \cite{blum} and on the
model approaches \cite{land,koch,golo}. Various forms have been
discussed including the extreme one $xg_1\sim \log^2 x$.
In this note we consider bound for $g_1$ by taking account of
unitarity in the direct channel. Corresponding constraint
for the transversity $h_1$ is given also.

For that purpose it is convinient to use the relation between
the distribution functions and the discontinuites in the helicity
amplitudes of the forward antiquark--hadron scattering \cite{jaffe}
which is based on the dominance of the ``handbag'' diagrams in DIS:
\begin{eqnarray}
q(x) & = & \frac{1}{2}\mbox{Im}[F_1(s,t)+F_3(s,t)]|_{t=0}\nonumber \\
\Delta q(x) & = & \frac{1}{2}\mbox{Im}[F_3(s,t)-F_1(s,t)]|_{t=0},\label{dq}
\end{eqnarray}
where $s\simeq Q^2/x$ and $F_i$ are the helicity amplitudes for the
elastic quark--hadron scattering in the notations for
the nucleon--nucleon scattering, i.e.
\[
F_1\equiv F_{1/2,1/2,1/2,1/2},
\, F_2\equiv F_{1/2,1/2,-1/2,-1/2}, \,
F_3\equiv F_{1/2,-1/2,1/2,-1/2},
\, F_4\equiv F_{1/2,-1/2,-1/2,1/2}
\]
and
\[ F_5\equiv F_{1/2,1/2,1/2,-1/2}.
\]
 We consider a quark--hadron scattering
similar to a hadron--hadron scattering. It can be justified
when considering quark as a structured hadronlike object. Arguments
in favour of such approach can be found in \cite{deld,bjor}.
We do not discuss here a possible influence of the off--mass--shell
effects.

Unitary representation for helicity amplitudes in the $U$--matrix
form of unitarization provides the following relations \cite{abk}
in the impact parameter representation:
\begin{equation}
 F_{\Lambda_1,\lambda_1,\Lambda_2,\lambda_2}(s,b)  =
U_{\Lambda_1,\lambda_1,\Lambda_2,\lambda_2}(s,b)+
i\rho (s)\sum_{\mu,\nu}
U_{\Lambda_1,\lambda_1,\mu,\nu}(s,b)
F_{\mu,\nu,\Lambda_2,\lambda_2}(s,b),\label{heq}
\end{equation}
where $\lambda_i$ and $\Lambda_i$ are the quark and hadron helicities,
respectively, and $b$ is the impact parameter.
The kinematical function $\rho(s)\simeq 1$ at $s\gg 4m^2$.
Explicit solution of Eqs. (\ref{heq}) has a rather complicated form:
\[
F_1(s,b)=\frac{\tilde{U}_1(s,b)[1-iU_1(s,b)]-i\tilde{U}_2(s,b)
U_2(s,b)}{[1-iU_1(s,b)]^2-[U_2(s,b)]^2},
\]
\[
F_3(s,b)=\frac{\tilde{U}_3(s,b)[1-iU_3(s,b)]-i\tilde{U}_4(s,b)
U_3(s,b)}{[1-iU_3(s,b)]^2-[U_4(s,b)]^2},
\]
where
\[
\tilde{U}_i(s,b)=U_i(s,b)+2U_5(s,b)F_5(s,b)
\]
and
\[
F_5(s,b) = \frac{U_5(s,b)}
{[1-iU_1(s,b)-iU_2(s,b)][1-iU_3(s,b)-iU_4(s,b)]-4U_5^2(s,b)}.
\]
 However, in the approximation when the helicity-flip functions
are much less than the helicity nonflip ones
one can get simple expressions
\begin{equation}
F_{1,3}(s,b)=\frac{U_{1,3}(s,b)}
{1-iU_{1,3}(s,b)}.\label{f13}
\end{equation}
The functions $F_i(s,t)$ are the corresponding Fourier--Bessel
transforms of the functions $F_i(s,b)$:
\begin{equation}
F_{1,3}(s,t)=\frac{s}{\pi^2}\int_0^\infty bdb F_{1,3}(s,b)
J_0(b\sqrt{-t}).\label{imp}
\end{equation}
Unitarity requires that Im$U_{1,3}(s,b)\geq 0$. Assume for simplicity
that the functions
$U_{1,3}(s,b)$ are pure imaginary, i.e.
$U_{1,3}(s,b)\rightarrow  iU_{1,3}(s,b)$, and parametrice these
functions in the form
\begin{eqnarray}
U_1(s,b) & = & \frac{1}{2}(1-a)U(s,b)\\
U_3(s,b) & = & \frac{1}{2}(1+a)U(s,b),
\end{eqnarray}
where the function $U(s,b)$ correspond to the unpolarized case
and $|a|\leq 1$. To maximaze the difference $U_3(s,b)-U_1(s,b)=aU(s,b)$
we consider $a$ as a constant.
For the function $U(s,b)$ we use simple form
\begin{equation}
U(s,b)=gs^\lambda e^{-\mu b}.\label{emb}
\end{equation}
This is a rather general parameterization for $U(s,b)$ which provides
correct analytical properties in the complex $t$--plane.
It follows also,
for example, from the chiral quark model for $U$--matrix \cite{csn}.
Note, that the following spectral representation is valid for
the function $U(s,b)$ \cite{spec}:
\begin{equation}
U(s,b)=\frac{\pi^2}{s}\int_{t_0}^\infty\rho(s,t)K_0(b\sqrt{ t})dt.
\label{spe}
\end{equation}
Another form of $U(s,b)$, e.g
\[
U(s,b)=gs^\lambda e^{-b^2/\omega(s)},\quad \omega(s)\sim \log s,
\]
which also leads to the total cross--section  saturating
the Froissart--Martin bound would provide the same results,
however, it does not respect analytical properties in the complex
$t$--plane. Indeed, as it follows from Eq. (\ref{spe}) the function
$U(s,b)$ should have a linear exponential dependence on the impact
parameter at large $b$.

Then using
Eq. (\ref{emb})
as an explicit form for $U(s,b)$
in Eqs. (\ref{f13}) and (\ref{imp})
it can be shown that Eqs. (\ref{dq}) provide at $x\to 0$
\begin{eqnarray}
\Delta q(x) & \sim & {a}/{(1-a^2)}{\log x}/{x}\quad
\mbox{for}\quad |a|\neq 1\nonumber\\
\Delta q(x) & \sim & {\log^2 x}/{x}\quad
\mbox{for}\quad |a|= 1
\end{eqnarray}
and
\begin{equation}
q(x)  \sim  {\log^2 x}/{x}.
\end{equation}
Since the above result are valid for each quark flavour
the similar behaviour
at $x\to 0$
will take place for the function $g_1(x)$.
Thus, unitarity together with parameterization
of the $U$--matrix which provides  saturation of the
Froissart--Martin bound  lead to the
following upper
bounds for the structure function $g_1(x)$:
\begin{equation}
g_1(x)\leq \log x/x\quad (|a|\neq 1)
\end{equation}
 and
\begin{equation}
g_1(x)\leq \log^2 x/x\quad (|a|=1).
\end{equation}
 The latter bound has been obtained earlier in
\cite{close}.

Using the relation between  the function $h_1^q(x)$ and corresponding
quark--hadron helicity amplitude $F_2(s,t)$ \cite{jaffe}:
\begin{equation}
h_1^q(x)=\frac{1}{2}\mbox{Im} F_2(s,t)|_{t=0} \label{h1}
\end{equation}
we can get similar upper bound for $h_1(x)$ at small $x$.
This function  measures the number density of quarks in
the transversity eigenstates
and is determined as a matrix element of the twist--two
transversity operator.

The unitary representation for the function $F_2(s,t)$
has the following form \cite{abk}:
\begin{equation}
F_2(s,t)=\frac{s}{\pi^2}\int_0^\infty bdb\frac{U_2(s,b)}{[1-iU_1(s,b)]^2}
J_0(b\sqrt{-t}).\label{f2}
\end{equation}

Using Eqs. (\ref{h1}),  (\ref{f2}) and proceeding through the same
steps as in the case of $g_1(x)$ we arrive to
the following extreme behaviour of $h_1(x)$ at $x\rightarrow 0$:
\begin{equation}
h_1(x)\sim\log{x}/x.
\end{equation}
Such behaviour of $h_1(x)$ could be considered,
in particular,
as an indirect confirmation of the inequality obtained
in \cite{jaffe}.

To have such upper bounds seems to be useful nowadays when the
experimental data indicate the possible rising behaviour of
$g_1(x)$ at small $x$ and the studies of $h_1(x)$
are planned at RHIC.

\vspace{0.5cm}

The author is grateful to N. E. Tyurin for  his helpful  comments
and discussions.

\end{document}